\documentclass[a4paper,11pt]{article}
\pdfoutput=1 

\usepackage{jcappub} 

\usepackage[T1]{fontenc}
\usepackage{bm}
\usepackage{latexsym}
\usepackage{dcolumn}
\usepackage{amsfonts,amssymb,amsmath}
\usepackage{graphicx,epsfig}
\usepackage{psfrag}
\usepackage{comment}
\usepackage{url}
\usepackage{feynmf}
\usepackage{rotating}
\usepackage{hyperref}
\usepackage{enumitem}
\usepackage{multirow}
\usepackage{ulem}
\usepackage[table]{xcolor}
\hypersetup{
     unicode=false,          
     pdftoolbar=true,        
     pdfmenubar=true,        
     pdffitwindow=false,     
     pdfstartview={FitH},    
     pdftitle={My title},    
     pdfauthor={Author},     
     pdfsubject={Subject},   
     pdfcreator={Creator},   
     pdfproducer={Producer}, 
     pdfkeywords={keyword1} {key2} {key3}, 
     pdfnewwindow=true,      
     colorlinks=true,        
     linkcolor=red,          
     citecolor=blue,         
     filecolor=magenta,      
     urlcolor=blue,         
     linktocpage=true
}

\def\beq{\begin{equation}}
\def\eeq{\end{equation}}
\def\br{\begin{eqnarray}}
\def\er{\end{eqnarray}}
\def\benu{\begin{enumerate}}
\def\efnu{\end{enumerate}}

\title{\boldmath A search for super-imposed oscillations to the primordial power spectrum in Planck and SPT-3G 2018 data}

\author[a,b]{A. Antony,}
\author[c,d]{F. Finelli,}
\author[e,f,c]{D. K. Hazra,}
\author[c,d]{D. Paoletti,}
\author[g,h]{A. Shafieloo}

\affiliation[a]{Asia Pacific Center for Theoretical Physics, Pohang, 37673, Republic of Korea}
\affiliation[b]{Department of Physics, POSTECH, Pohang, 37673, Republic of Korea}
\affiliation[c]{INAF/OAS Bologna, Osservatorio di Astrofisica e Scienza dello Spazio, Area della ricerca CNR-INAF, via Gobetti 101, I-40129 Bologna, Italy}
\affiliation[d]{INFN,  Sezione  di  Bologna,  via  Irnerio  46,  40126  Bologna,  Italy}
\affiliation[e]{The Institute of Mathematical Sciences, HBNI, CIT Campus, Chennai 600113, India}
\affiliation[f]{Homi Bhabha National Institute, Training School Complex, Anushakti Nagar, Mumbai 400085, India}
\affiliation[g]{Korea Astronomy  and Space Science Institute, Daejeon 34055, Korea}
\affiliation[h]{Korea University of Science and Technology, Daejeon 34113, Korea}

\emailAdd{akhil.antony@apctp.org}
\emailAdd{fabio.finelli@inaf.it}
\emailAdd{dhiraj@imsc.res.in}
\emailAdd{daniela.paoletti@inaf.it}
\emailAdd{shafieloo@kasi.re.kr}

\abstract{
We search for super-imposed oscillations linearly or logarithmically spaced in Fourier wavenumbers $k$ in Planck and South Pole Telescope (SPT-3G) 2018 temperature and polarization data. The SPT-3G temperature
and polarization data provide a new window to test these oscillations at high multipoles beyond
the Planck angular resolution and sensitivity. We consider linear and logarithmic oscillations with a constant amplitude, or with a power-law dependence or a Gaussian modulation, always in $k$. These models correspond to three, four and five additional parameters beyond power-law primordial power spectrum for the templates considered, respectively. We find that each of the five models considered can
provide an improved fit to Planck data, consistently with previous findings, and to SPT-3G data, always compared to power-law power spectrum. {We find tighter constraints on the amplitude of the super-imposed oscillations from the combined Planck/SPT-3G data set than in each individual data sets. For linear oscillations, with the amplitude allowed to vary as a power-law in $k$, as in the case of EFT, we find that the addition of SPT-3G data sets tighter constraints on the possibility that the amplitude increase at small scales.}   
When the ranges of parameters which provide
a better fit to Planck and SPT-3G data overlap, as in the case of Gaussian modulated oscillations, we find 
a larger $\Delta \chi^2 \sim - 17.5 \, (-14.7)$ for logarithmic (linear) oscillations - in a combined Planck/SPT-3G data set than 
in each individual data sets.
These findings will be further tested with upcoming CMB temperature and polarization
measurements at high multipoles provided by ongoing ground experiments.
}

\begin{document}
\maketitle
\flushbottom

\section{Introduction}
The power spectrum of primordial scalar fluctuations is being tested by multiple cosmological probes
at different scales and redshifts. A power law spectrum embedded in the six parameters $\Lambda$CDM model with power law primordial spectrum
is a good fit to Planck cosmic microwave background (CMB) anisotropy data, which currently has the
lion's share of the constraining power among different cosmological observables within the range $ \sim [0.001,\,0.1] \,{\rm Mpc} ^{-1}$.
However, there are certain deviations from a power law spectrum which are still
allowed by the current precision and angular resolution of Planck 2018 data \cite{Planck:2018inf}. Some of these deviations lead to an improved fit with a $\Delta \chi^2 \sim -[10-15]$ compared to a power-law spectrum, although the extra parameters
beyond the amplitude $A_\mathrm{s}$ and the spectral index $n_\mathrm{s}$ penalize these models from a Bayesian point of view \cite{Planck:2018inf}.

Among the deviations from a power-law in the primordial spectrum, super-imposed oscillations, either linear or logarithmic in wavenumber, stand out due to their deeper theoretical foundations and improved fit to the Planck data.
Non-vacuum initial vacuum states for quantum fluctuations \cite{Kempf:2000ac, Easther:2001fz, Martin:2003kp}, resonant models including periodic oscillations
in the potential \cite{Chen:Osc000} as axion monodromy \cite{Flauger:Osc1} are among the mechanisms which generate super-imposed logarithmic oscillations, whereas sharp turns generate super-imposed linear oscillations \cite{Jackson:2013vka}. 
These models have therefore become a workhorse for
exploration with CMB anisotropy data \cite{Planck:2018inf,Easther:2004vq, Planck:2015sxf,Planck:2013jfk,Meerburg:2013cla,Meerburg:2013dla,Peiris:2013opa,Easther:2013kla, Aich:Osc3,Hamann:2021eyw},
also by a joint power-spectrum/bispectrum analysis \cite{Fergusson:2014tza,Meerburg:2015owa}, and 
with galaxy surveys data \cite{Hazra:2012vs, Beutler:2019ojk, Ballardini:2022vzh, Mergulhao:2023ukp}. These models constitute a benchmark
for future CMB experiments \cite{CORE:2016ymi,Hazra:2017WWI}, as well as for future and ongoing galaxy surveys
\cite{Huang:2012mr, Ballardini:2016hpi, Chen:2016vvw, Chen:2016zuu, Chen:2020ckc, Debono:2020emh,Euclid:2023shr}. See \cite{Covi:2006ci,Hamann:2007pa,Hazra:2010ve,PhysRevD.70.043523,Achucarro:2013cva,Hazra:2014jwa,Chen:2014SC,Nicholson:2009pi,Hunt:2015iua,Braglia:2021SC1,Braglia:2021sun,Antony:2021bgp, Hazra:2022rdl,Antony:2022ert} for other searches of primordial features in CMB data.

CMB polarization anisotropies should be a refined test for primordial features,
given the sharpness of the polarization transfer
functions compared to those in temperature \cite{Chluba:2015Step}. Whereas $Planck$ remains the main full sky data set until LiteBIRD
\cite{LiteBIRD:2022cnt}, better data at higher resolution from ground experiments such as Atacama Cosmology Telescope (ACT) \cite{ACT:2020gnv} and South Pole Telescope (SPT) \cite{SPT-3G:2022hvq} are already in the public domain. The high degree of consistency between $Planck$ and SPT-3G data has been discussed in \cite{SPT-3G:2022hvq}.
We therefore explore oscillations linear and logarithmically spaced in $k$ with a constant and a Gaussian modulated amplitude with $Planck$ \cite{Planck:2019Like} and, for the first time, with SPT-3G 2018 data \cite{SPT-3G:2022hvq}, as well with their combination
\footnote{See \cite{Hazra:2022rdl} for a study of different primordial features with $Planck$ and ACT DR4 data \cite{ACT:2020gnv}. Note however that $Planck$ and ACT DR4 data lead to discrepant estimates for $\Omega_b h^2$ and $n_\mathrm{s}$ \cite{ACT:2020gnv}.}.

This paper is structured as follows. After the introduction, in Section 2 we describe the four models chosen as representative for super-imposed oscillations, the data sets used and the sampling methodology. In Section 3, we describe our results and we conclude in Section 4.
    
\section{Methodology}~\label{sec:data}
\begin{figure*}[!hb]
\centering
\includegraphics[width=\textwidth]{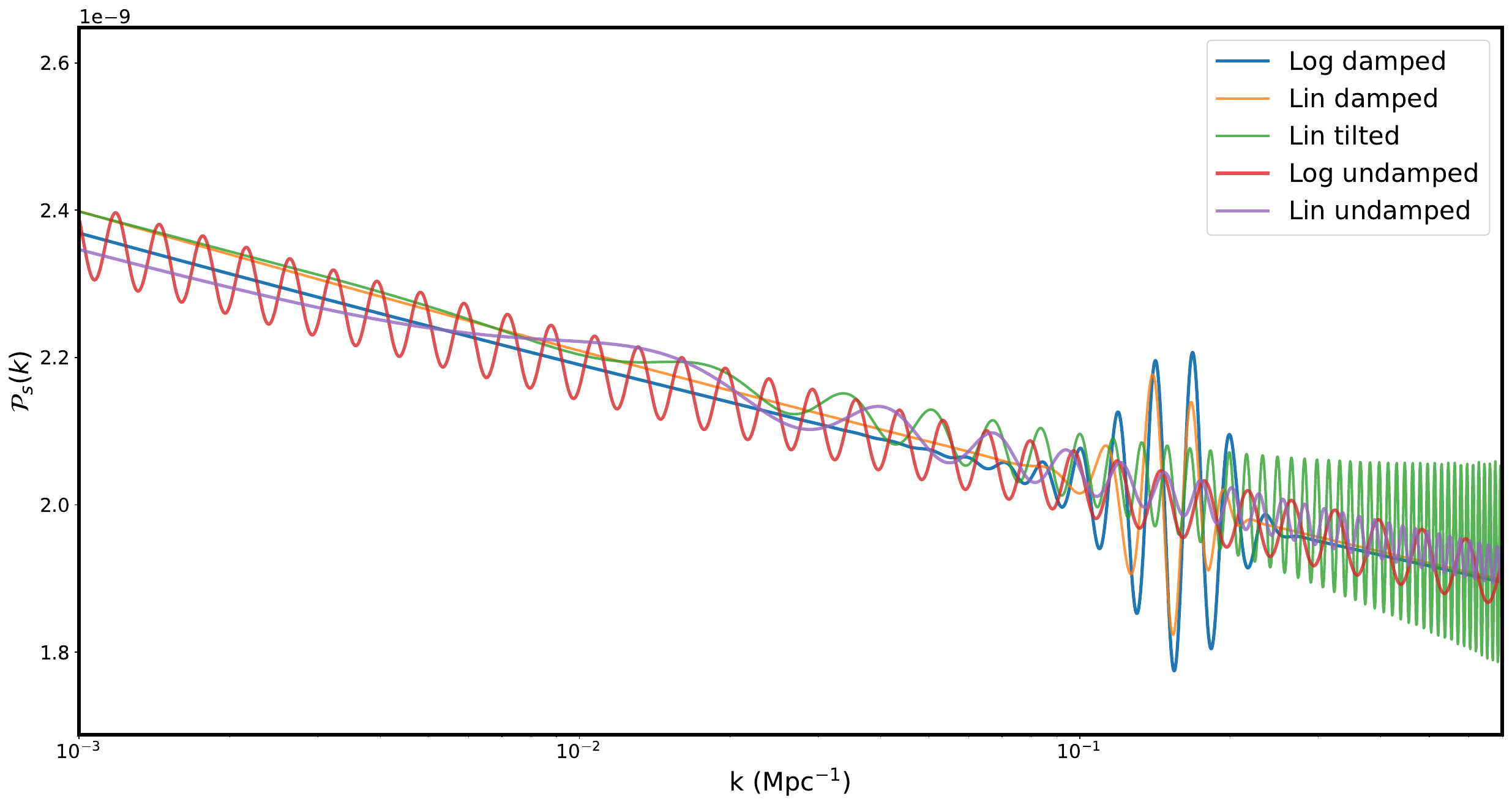}
\caption{\label{fig:curvature_powerspectra} Primordial curvature power spectra for representative models of the  {five} templates considered. 
}
\end{figure*}

In this paper, we investigate the effect of sinusoidal oscillations in the primordial power spectrum which are periodic on linear or logarithmic scales. In our study, we further examine two class of models: one where the oscillations span the entire observable range of wavenumbers ($k$), and another where the oscillations are localized. In the latter case, these oscillations undergo damping on both sides of a central peak. In total, we analyse four models in this study: linear and logarithmic oscillations, with constant amplitude and Gaussian modulated amplitude, as can be seen in Fig. \ref{fig:curvature_powerspectra} 

\subsection{Models and Priors}
We consider the following templates for global oscillations:

\begin{equation}
    P_{\rm global}^{\rm lin}(k) = P_0(k) \left[  1 + \alpha {\left(\frac{k}{k_*}\right)^{n_\mathrm{lin}}}\cos \left( \omega \, \frac{k}{k_*}+\phi \right) \right]  \label{eq:undamped-lin}
\end{equation}
\begin{equation}
    P_{\rm global}^{\rm log}(k) = P_0(k) \left[ 1 +
    \alpha\cos \left( \omega \ln \frac{k}{k_*} + \phi \right) \right] \label{eq:undamped-log}
\end{equation}
These oscillations have an amplitude $\alpha$, frequency $\omega$, a phase $\phi$ and {a tilt $n_\mathrm{lin}$ for the additional power-law $k$-modulation for linear oscillations \cite{Jackson:2013vka,Meerburg:2013cla,Meerburg:2013dla, Planck:2015sxf},}
with priors listed in Table~\ref{prior_table_features}. The priors are the same of those employed in Ref. \cite{Planck:2018inf}.
Here $P_0(k)$ is the power-law power spectrum given as $A_s(k/k_*)^{n_s-1}$ where $k_*=0.05 \,{\rm Mpc} ^{-1}$.

As localized oscillation models extending Eqs. ~\ref{eq:undamped-lin},~\ref{eq:undamped-log}, we consider :
\begin{equation}
    P_{\rm local}^{\rm lin}(k) = P_0(k)\bigg[ 1 + \alpha\cos\left(\omega \, \frac{k}{k_*}+\phi \right) e^{-\frac{\beta^2 (k-\mu)^2}{2 k_*^2}}  \bigg] \label{eq:damped-lin}
\end{equation}

\begin{equation}
P_{\rm local}^{\rm log}(k) = P_0(k)\bigg[ 1 + \alpha\cos \left( \omega \ln \frac{k}{k_*} + \phi \right) e^{-\frac{\beta^2 (k-\mu)^2}{2 k_*^2}}  \bigg] \label{eq:damped-log}
\end{equation}
that contain five parameters, two, i.e. the center $\mu$ and the width $\beta$ of the Gaussian envelope, in addition to our global oscillation model.
In this work we consider varying all the five parameters with priors as in Table~\ref{prior_table_features}. We have explicitly checked that the dimensional reduction adopted in \cite{Hazra:2022rdl}, by fixing the phase in such a way that the peak of the Gaussian is located at a maximum of the oscillations $\phi = - \omega \mu/k_*$ ($\phi = - \omega \ln (\mu/k_*)$) for the linear (logarithmic) case, modifies the results in the following way. The distribution of $\chi^2$ and the marginalized posterior for the amplitude are quite robust to fixing the phase. The Bayesian ratio, instead, would be different because of the smaller prior volume when fixing the phase.
The prior ranges for these five parameters are listed in Table~\ref{prior_table_features}. Note that the priors chosen for $\alpha \,, \omega \,, \phi$ are the same as those of of the global oscillations case.

\begin{table}[h]
\centering

\begin{tabular}{|l|c|c|c|c|c|c|} 
\hline
                & $\alpha$        & $\log_{10} \omega$ & $n_\mathrm{lin}$        &$\mu$                 & $\beta$       & $\phi/(2\pi)$ \\ \hline
Linear & {[}0, 0.5{]} & {[}0,2.1{]}   & $n_\mathrm{lin}=0$ &      -            &     -       & {[}0,1{]} \\ \hline
Log    & {[}0, 0.5{]} & {[}0,2.1{]}   &  - &       -           &       -     & {[}0,1{]} \\ \hline
{Linear tilted} & {[}0, 0.5{]} & {[}0,2.1{]}   &   {[}-1,1{]}   &       -           &       -     & {[}0,1{]} \\ \hline
Damped Linear   & {[}0, 0.5{]} 
& {[}0,2.1{]} & - 
& {[}0.001, 0.175{]} & {[}0,10{]} &    {[}0,1{]}     \\ \hline
Damped Log      & {[}0, 0.5{]} & {[}0,2.1{]} & - &  {[}0.001, 0.175{]} & {[}0,10{]} &    {[}0,1{]}      \\ \hline
\end{tabular}
\caption{Prior table for parameters of the super-imposed oscillation templates.  
\label{prior_table_features}}
\end{table}

The priors for the six remaining cosmological parameters are the standard priors used in Polychord~\cite{Handley:Polychord} and are given in Table~\ref{prior_table_parameters}. These priors are used for each of the four templates.

\begin{table}[h]
\centering

\begin{tabular}{|l|c|c|c|c|c|c|} 
\hline       
Parameter & $\Omega_{\mathrm b} h^2$        & $\Omega_{\mathrm c} h^2$         & $100 \theta_{\mathrm MC}$                 & $\tau$       & $n_{\mathrm s}$ & $\log[10^{10} A_{\mathrm s}]$ \\ \hline
Prior & {[}0.019, 0.025{]} & {[}0.095,0.145{]}   & {[}1.03,1.05{]}  &    {[}0.01,0.4{]}     & {[}0.885,1.04{]} & {[}2.5,3.7{]} \\ \hline
\end{tabular}
\caption{Prior table for cosmological parameters. Note that the priors on the background parameters are narrower than the standard MCMC analyses performed with $Planck$ data. For nested sampling we use this restricted prior range 
throughout our analyses as suggested in the PolyChord sampler~\cite{handley_cosmochord}}.
\label{prior_table_parameters}
\end{table}

\subsection{Datasets and Samplers}

Our analyses in this paper are based only on CMB data. We work with 3 combinations of datasets, namely $Planck$ (P18TP), SPT-3G (SPT3G) and combined P18TP and SPT-3G (P18TP+SPT3G). 

We use $Planck$ 2018 release temperature and polarization likelihoods \cite{Planck:2019Like}. Following the primordial feature analysis of $Planck$~\cite{Planck:2018inf}, we do not consider $Planck$ lensing likelihood. 
Differently from the $Planck$ analysis for features~\cite{Planck:2018inf}, we restrict ourselves to binned data, 
 but all foreground and calibration nuisance parameters are allowed to vary along with their recommended priors \cite{Planck:2019Like}. In this paper we consider Plik binned data since the SPT-3G data are provided in a binned format, inhibiting the joint exploration of the high frequency oscillations domain allowed by unbinned Planck data.

For SPT-3G 2018 data, we use temperature, polarization auto and cross correlation data from 90, 150 and 220 GHz channels~\footnote{https://pole.uchicago.edu/public/data/balkenhol22/} \cite{SPT-3G:2022hvq}. When we use SPT only likelihood for analysis, in order to break the degeneracy between the  optical depth and the amplitude of primordial fluctuations, we use the optical depth prior from $Planck$ 2018 baseline analysis, $\tau=0.054\pm0.0074$, as recommended in \cite{SPT-3G:2022hvq}. Here, too, we allow the default nuisance parameters to vary.

\begin{figure*}[!htb]
\centering
\includegraphics[width=\textwidth]{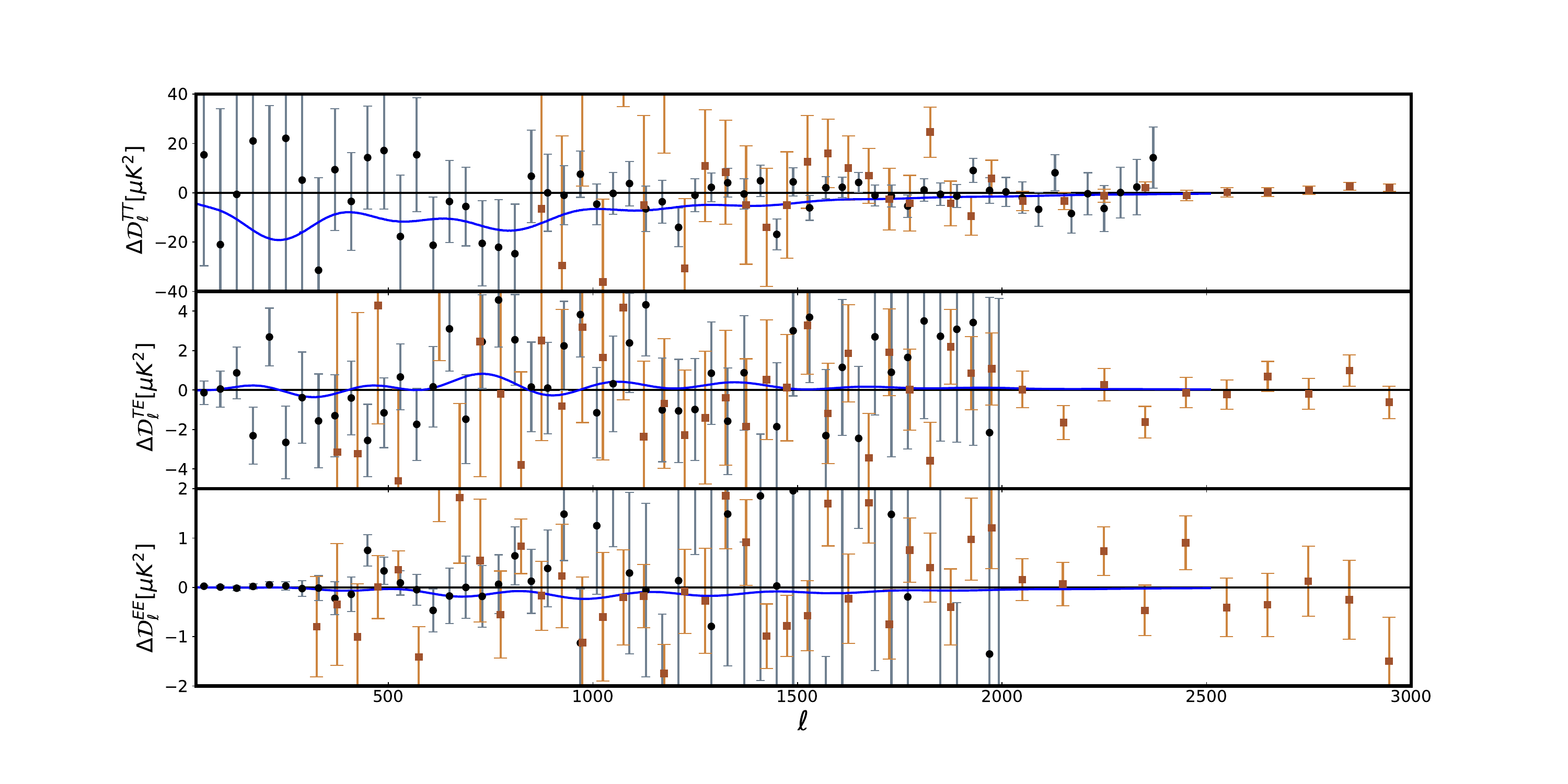}
\caption{\label{fig:P18TP_vs_SPT} 
$Planck$ (black) and SPT-3G TT, TE, EE (orange) binned residuals to the $Planck$ 2018 TP power law best-fit. The relative difference of the power law bestfit to SPT-3G data plus $Planck$ 2018 measurement of $\tau$ with respect to the $Planck$ 2018 TP power law bestfit is also shown for comparison (blue line).}
\end{figure*}

The consistency and complementarity of $Planck$ 2018 and SPT-3G 2018 data is shown in Fig.~\ref{fig:P18TP_vs_SPT}.
Given the lack of publicly available covariance between $Planck$ and SPT-3G data, we consider as baseline a conservative combined $Planck$/SPT-3G data set in which there is no-overlap in multipoles between the two datasets. In gluing the two data sets, we follow indications in the SPT-3G release~\cite{SPT-3G:2022hvq}, where it has been discussed that SPT-3G data have smaller uncertainties than $Planck$ above multipole $\ell=2000$ for TT, $\ell=1400$ for TE and $\ell=1000$ for EE. Following that, we use a truncated version of $Planck$ Plik TTTEEE likelihood with TT in the range $\ell=30-2000$, TE in $\ell=30-1400$ and EE in $\ell=30-1000$. Therefore, in this combined data set, we use TT from $\ell=2000$, TE from $\ell=1400$ and EE from $\ell=1000$ for SPT-3G data. Note that with this conservative choice, the multipole range $10^3 \lesssim \ell \lesssim 2 \times 10^3$ are tested by $Planck$ in TT and by SPT-3G in EE. 

We do not use ACT DR4 \cite{ACT:2020gnv} due to the observed 3$\sigma$ tension with $Planck$. Recently, a couple of authors in this paper has demonstrated~\cite{Hazra:2024nav} that the disagreement in the primordial spectrum between $Planck$ PR3 and ACT DR4 exists within $0.08-0.16$ (1/Mpc). Since this paper discusses the primordial power spectrum constraints, a dataset that is in tension with $Planck$ if included, can introduce bias in the parameter estimation. Therefore, we decided to work with $Planck$ PR3 and SPT3G datasets in a conservative, non-overlapping combination. A subset of authors of this paper had also used $Planck$ and ACT in constraining the damped sin oscillation~\cite{Hazra:2022rdl} in the context of solving the lensing anomalies and cosmological tensions together. In that paper, the analysis was performed within a prior range restricted around a characteristic frequency of oscillation. Since within that range for a particular model, the $Planck$ data prefers higher spectral tilt  that reduces tension with the ACT data considerably, a joint analysis could be justified.

Given the multi-modal nature of posteriors, Markov Chain Monte Carlo (MCMC) is not expected to be effective in sampling the
parameter space. We therefore use the Polychord~\cite{Handley:Polychord} sampler for nested sampling of parameter spaces and for the computation of Bayes factors. 
We use $n_{\rm live}=2048$ for all analyses ($n_{\rm live}=2000$ was also used in Ref. \cite{Planck:2018inf} for the search of features).

In settling for live points related to combined analysis, we compared MCMC results with Polychord for a power-law spectrum.
For $n_{\rm live}=2048$ parameter posteriors obtained from Polychord converge to MCMC posteriors. We noticed that for the feature models too, $n_{\rm live}=2048$
is a conservative choice for the convergence of the posteriors.
In obtaining best-fits, we use BOBYQA minimizer \cite{BOBYQA}. We identify the samples from the Polychord chains corresponding to maximum likelihood peak and we obtain local minima by BOBYQA around that peak.

\begin{table}[]
\begin{tabular}{|l|ll|ll|ll|}
\hline
 Models  & \multicolumn{1}{l} {$\Delta\chi^2_\mathrm{P18}$} & $\ln B_\mathrm{P18}$ & \multicolumn{1}{l}{ $\Delta\chi^2_\mathrm{SPT}$} & $\ln B_\mathrm{SPT}$ & \multicolumn{1}{l} {$\Delta\chi^2_\mathrm{P18+SPT}$} & $\ln B_\mathrm{P18+SPT}$ \\ \hline 
Lin undamped      & \multicolumn{1}{l|}{-11.8} & -2.6 $\pm$ 0.3    & \multicolumn{1}{l|}{-7.0}  & -1.8 $\pm$ 0.3    & \multicolumn{1}{l|}{-12.0}  & -4.5 $\pm$ 0.4     \\ \hline
Log undamped      & \multicolumn{1}{l|}{-9.3} & -2.9 $\pm$ 0.3     & \multicolumn{1}{l|}{-12.0} & -1.2 $\pm$ 0.3     & \multicolumn{1}{l|}{-14.3}  & -6.0 $\pm$ 0.4     \\ \hline
Lin tilted      & \multicolumn{1}{l|}{-11.8} & -2.7 $\pm$ 0.3   & \multicolumn{1}{l|}{-9.1}  & -1.5 $\pm$ 0.3    & \multicolumn{1}{l|}{-14.5}  & -3.1 $\pm$ 0.4    \\ \hline 
Lin damped        & \multicolumn{1}{l|}{-11.8} & -2.0 $\pm$ 0.3     & \multicolumn{1}{l|}{-7.7}  & 0.0 $\pm$ 0.3     & \multicolumn{1}{l|}{-14.7} & -1.8 $\pm$ 0.3      \\ \hline
Log damped        & \multicolumn{1}{l|}{-10.0}  & -1.8  $\pm$ 0.3     & \multicolumn{1}{l|}{-12.0} & -0.4 $\pm$ 0.3     & \multicolumn{1}{l|}{-17.5} & -3.1 $\pm$ 0.4     \\ \hline
\end{tabular}
\caption{
Improvement in fit to the data and Bayesian evidence obtained by the templates with respect to the power law primordial power spectrum for the three data sets considered. We have studied the templates, both independently and jointly, against  $Planck$ and  SPT dataset. In all cases we get a good improvement in fit to the data with respect to power law power spectrum, although none of the models are preferred in terms of Bayesian evidence because of the number of additional parameters.  
\label{results_table}}
\end{table}

\begin{figure*}[!h]
\centering
\includegraphics[width=0.48\textwidth]{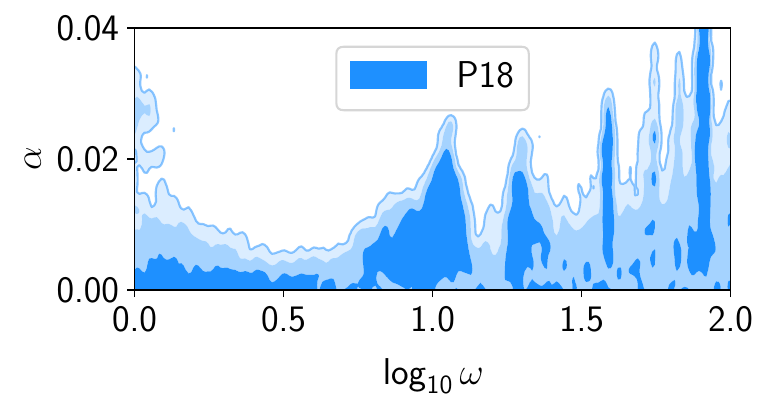}
\includegraphics[width=0.48\textwidth]{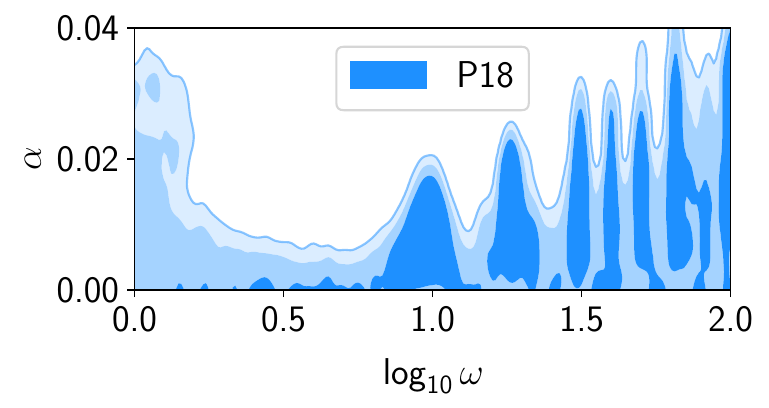}
\caption{\label{fig:P18TP_2D} 
Marginalized joint 68 \%, 95 \% and 99 \% regions for $(\alpha, \omega)$ for linear (left) and logarithmic undamped wiggles. Our results are quite consistent with those in Ref. \cite{Planck:2018inf}, which are instead obtained with unbinned data and fixed nuisance/foreground parameters. 
}
\end{figure*}
\vspace{1cm}

\begin{figure*}[!h]
\centering
\includegraphics[width=0.48\textwidth]{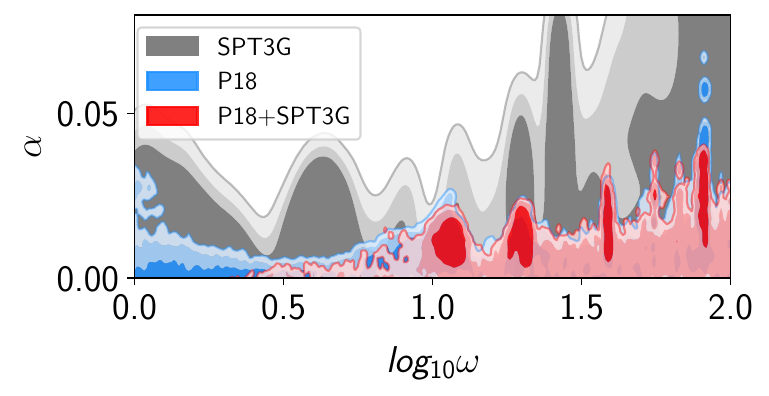}
\includegraphics[width=0.48\textwidth]{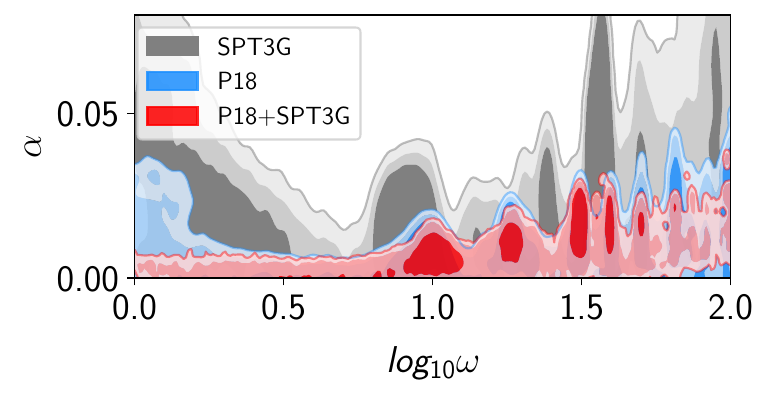}
\caption{\label{fig:undamped_2D} 
Marginalized joint 68 \% and 95 \% regions for $(\alpha, \omega)$ for linear (left) and logarithmic undamped wiggles with $Planck$ (blue), SPT-3G (grey) and $Planck$ plus SPT-3G (red) data.   
}
\end{figure*}
\vspace{1cm}

\begin{figure*}[!h]
\centering
\includegraphics[width=0.48\textwidth]{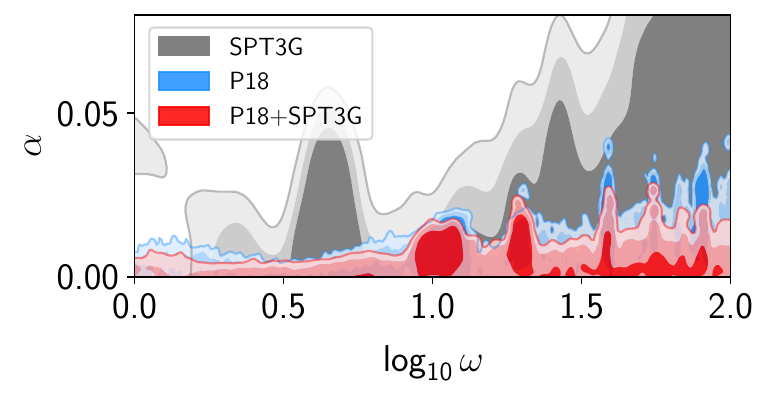}
\includegraphics[width=0.48\textwidth]{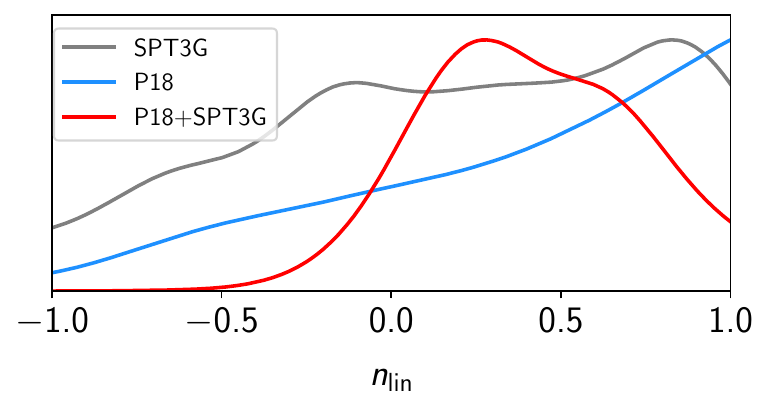}
\caption{\label{fig:undampedtilted_2D} 
Left panel: marginalized joint 68 \% and 95 \% regions for $(\alpha, \omega)$ for linear tilted super-imposed oscillations. Right panel: marginalized contour for $n_\mathrm{lin}$. As for Fig. 4, $Planck$ is in blue, SPT-3G in grey and $Planck$ plus SPT-3G in red.  
}
\end{figure*}
\vspace{1cm}

\begin{figure*}[!h]
\centering
\includegraphics[width=0.48\textwidth]{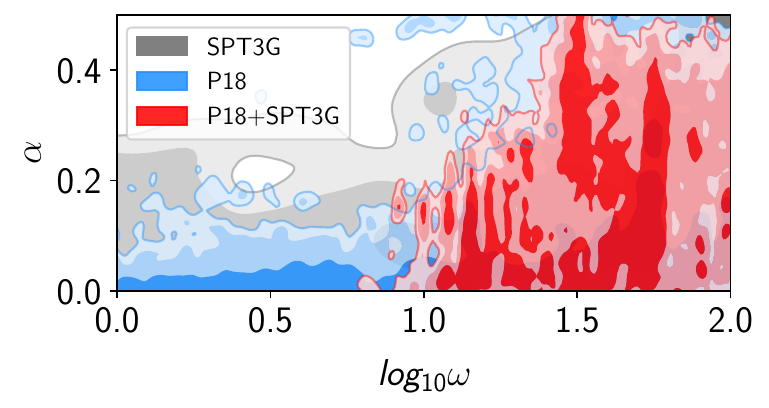}
\includegraphics[width=0.48\textwidth]{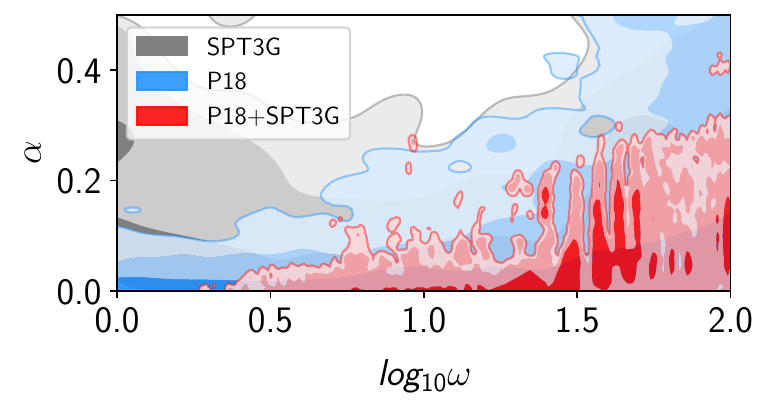}
\caption{\label{fig:damped_2D} 
Marginalized joint 68 \% and 95 \% regions for $(\log_{10} \omega \,, \alpha)$ for linear (left) and log
}
\end{figure*}
\vspace{1cm}


\section{Results}

Results for $Planck$, SPT-3G and the baseline combination of $Planck$ and SPT-3G data are presented in Table \ref{results_table}. 

We first discuss the results obtained for the undamped templates as given in Eqs.~\ref{eq:undamped-lin},~\ref{eq:undamped-log}.
We find a very good agreement with the results in \cite{Planck:2018inf} when using $Planck$ 2018 unbinned data and nuisance/foreground parameters fixed to the power law bestfits. 
The joint regions we obtain in Fig. \ref{fig:P18TP_2D} are quite similar in shape to those in Ref. \cite{Planck:2018inf}, but broader because we use a binned likelihood and the nuisance/foreground parameters are allowed to vary. Linear and logarithmic undamped templates lead to an improvement $\Delta \chi^2_\mathrm{P18} \sim - 11.8$ and $\sim - 9.3$, respectively, compatible with previous results \cite{Planck:2018inf,Braglia:2021rej}. 

With SPT-3G 2018 and the $Planck$ prior on $\tau$, we find a prominent peak in the posterior for $\log_{10}\omega \sim 1.5$ with $\Delta \chi^2_\mathrm{SPT} \sim - 12.0$ for logarithmic oscillations, whereas we find a smaller improvement in the fit for linear oscillations, always compared to power-law power spectrum. For SPT-3G data, we find that a contribution to this improvement comes from fitting logarithmic oscillations in the range $1000 \lesssim \ell \lesssim 2000$.

For these global oscillation models, the posteriors peak at different frequency for $Planck$ and SPT-3G 2018 data. 
For the $Planck$ + SPT3G data set, we find a $\chi^2_\mathrm{P18+SPT} \sim - 12 \, (-14.3) $ for linear (logarithmic) undamped oscillations. 
As is clear from Fig.~\ref{fig:undamped_2D}, this combined $Planck$ + SPT-3G data set constrains the amplitude of these oscillations in a tighter way compared to the individual data sets. We obtain $\alpha \lesssim 0.031 $ ($\alpha \lesssim 0.023$) at 95 \% CL from $Planck$ plus SPT-3G for linear (logarithmic) oscillations, which improve on the corresponding $Planck$ 95 \% CL constraint $\alpha \lesssim 0.036 $ ($\alpha \lesssim 0.031 $). 
{When the amplitude for superimposed linear oscillations is allowed to vary as a power-law in $k$ as in \cite{Jackson:2013vka} we find that the SPT-3G data are effective in constraining $n_\mathrm{lin}$, see Fig. \ref{fig:undampedtilted_2D}. In particular, we obtain $n_\mathrm{lin} = 0.38 \pm 0.30 $ at 68 \% CL with $Planck$ combined with SPT-3G data, whereas we do not find any meaningful constraint when the two datasets are considered separately.} 
None of {these three} models are statistically preferred over the power law model for any data set considered. 

We analyzed the damped templates in Eqs.~\ref{eq:damped-lin},~\ref{eq:damped-log}. SPT-3G data can constrain the envelope of these templates at scales beyond the $Planck$ resolution as the one mimicking the $A_{\rm L}$ effect.
The $(\log_{10} \omega \,, \alpha)$ plots for the damped linear and logarithmic template corresponding to all three datasets are provided in Fig.~\ref{fig:damped_2D}. 
We find that the Gaussian modulated linear oscillations provide a larger improvement in the fit to $Planck$ 
data ($\Delta\chi^2_\mathrm{P18} \sim -11.8$) than Gaussian modulated logarithmic oscillations ($\Delta\chi^2_\mathrm{P18} \sim -10$). For SPT-3G data, the opposite occurs, i.e. Gaussian modulated logarithmic oscillations provide a better fit compared to linear oscillations ($-12.0$ vs $-7.7$).
For a Gaussian modulated amplitude, we find a prominent peak at $\log_{10} \omega \sim 1.55$ 
for the SPT-3G data set (combined with the prior on $\tau$ from Planck 2018). 
These oscillations are related, but do not correspond exactly, to oscillations in SPT-3G EE data found by Gaussian regression in \cite{Calderon:2023obf}\footnote{Oscillations in $C_\ell^{EE}$ are not necessarily fit by super-imposed oscillations
to the curvature power spectrum which are also imprinted in $C_\ell^{TT}$.}. 
Since the ranges of parameters which provide a better fit independently to $Planck$ and SPT-3G data approximately overlap for modulated oscillations, we find a larger $\Delta \chi^2$ in the combined $Planck$ plus SPT-3G data set, up to the maximum $\Delta \chi^2 \sim - 17.5 (- 14.7)$ for modulated logarithmic (linear) oscillations. It is interesting to note that SPT-3G data do not prefer $A_L$ larger than 1 \cite{SPT-3G:2022hvq}, but does not disfavour the modulated linear oscillation template which mimics the $A_L$ effect \cite{Planck:2018inf}.

Note that super-imposed oscillations with a Gaussian envelope have been studied with Planck data in the context of discordances in cosmological parameters such as $A_L$ \cite{Planck:2018inf} and $H_0$ \cite{ HazraHST:2018,Antony:2022ert}. These models showed improved fits to the Planck temperature and polarization data compared to the standard power-law model, while also yielding a higher $H_0$ and a lower value of $S_8$, in particular when the SH0ES measurement is inserted in the fit. Our analysis shows for the first time that the more precise combination with SPT-3G data does not decrease the improvement of these templates in the fit obtained with Planck data only. This is an important result which will be again verified by the higher sensitivity of future polarization data. On the other side, since we are adopting wide priors for primary parameters and we are not using SH0ES data in this paper, the marginalized posterior of $H_0$ does not show significant shift relative to the analysis with a standard power-law primordial spectrum.

We would like to note that while the P18+SPT-3G 2D marginalized contours are tighter than both P18 and SPT-3G constraints, the constraints on the amplitude get worse compared to P18. This degradation can be explained by the positive correlation $\alpha-\mu$. A higher value of amplitude of super-imposed oscillations is allowed when the feature is located at very small scales $\mu \gtrsim 0.1~\rm{Mpc}^{-1}$. In the joint likelihood, scales smaller than this wavenumber are not well constrained by $Planck$. Therefore, majority of the samples at $\mu \gtrsim 0.1~\rm{Mpc}^{-1}$ fit the SPT3G data better with the super-imposed oscillations while within the $Planck$ observed scales, the spectrum stays power law. For undamped cases, oscillations are present at all cosmological scales, we find that the joint dataset constrains the amplitude better than any individual datasets.

\begin{figure*}[!htb]
\centering
\includegraphics[width=\textwidth]{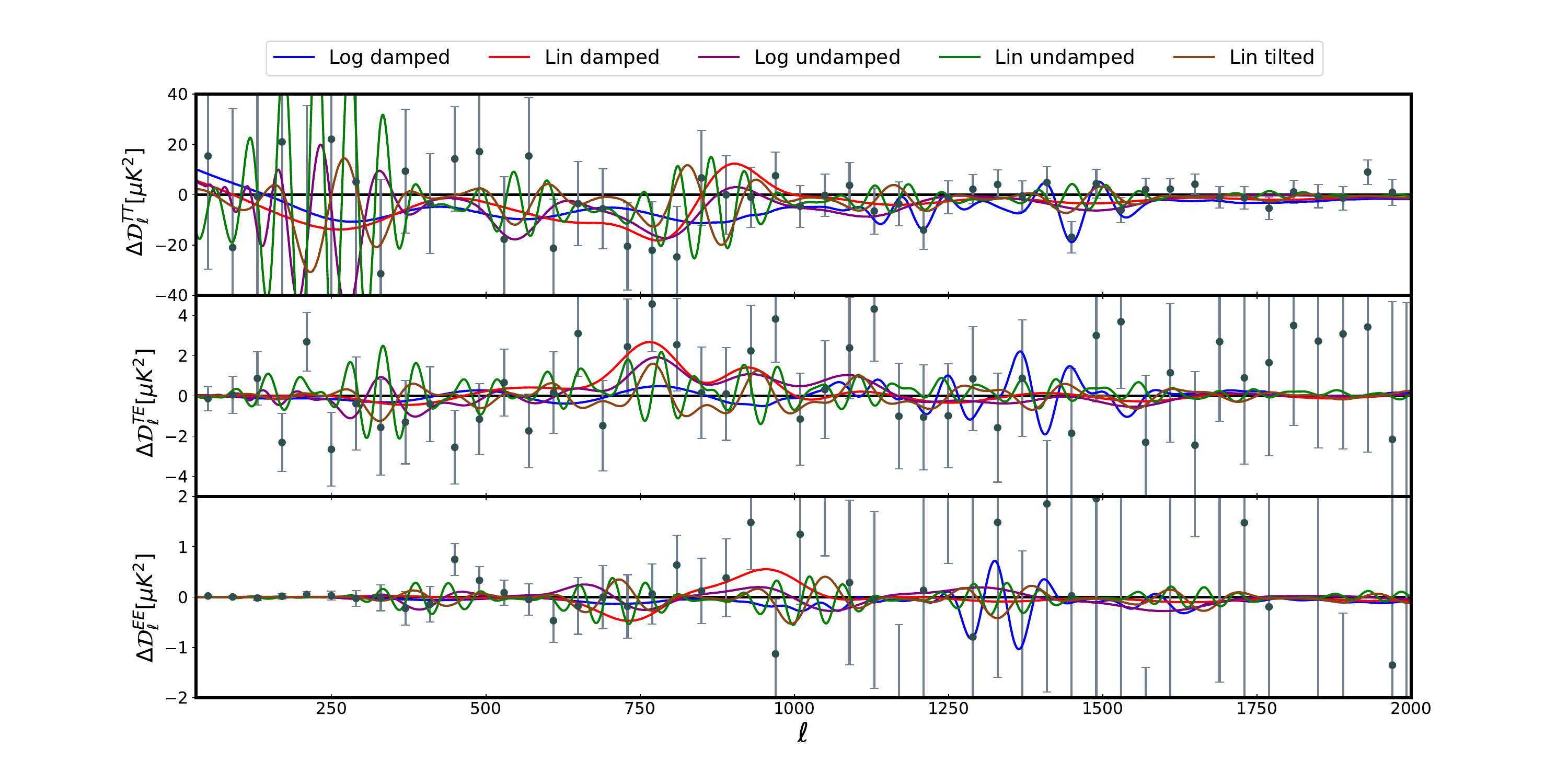} 
\caption{\label{fig:Residual_P18TP} Relative differences of the $Planck$ 2018 TP best fits for the four models considered with respect to the corresponding $Planck$ 2018 TP power law bestfit (see legends for the colors). Plik TT, TE, EE binned residuals (black data points) to the $Planck$ 2018 TP power law best-fit are also plotted for comparison. 
}
\end{figure*}

\begin{figure*}[!htb]
\centering
\includegraphics[width=\textwidth]{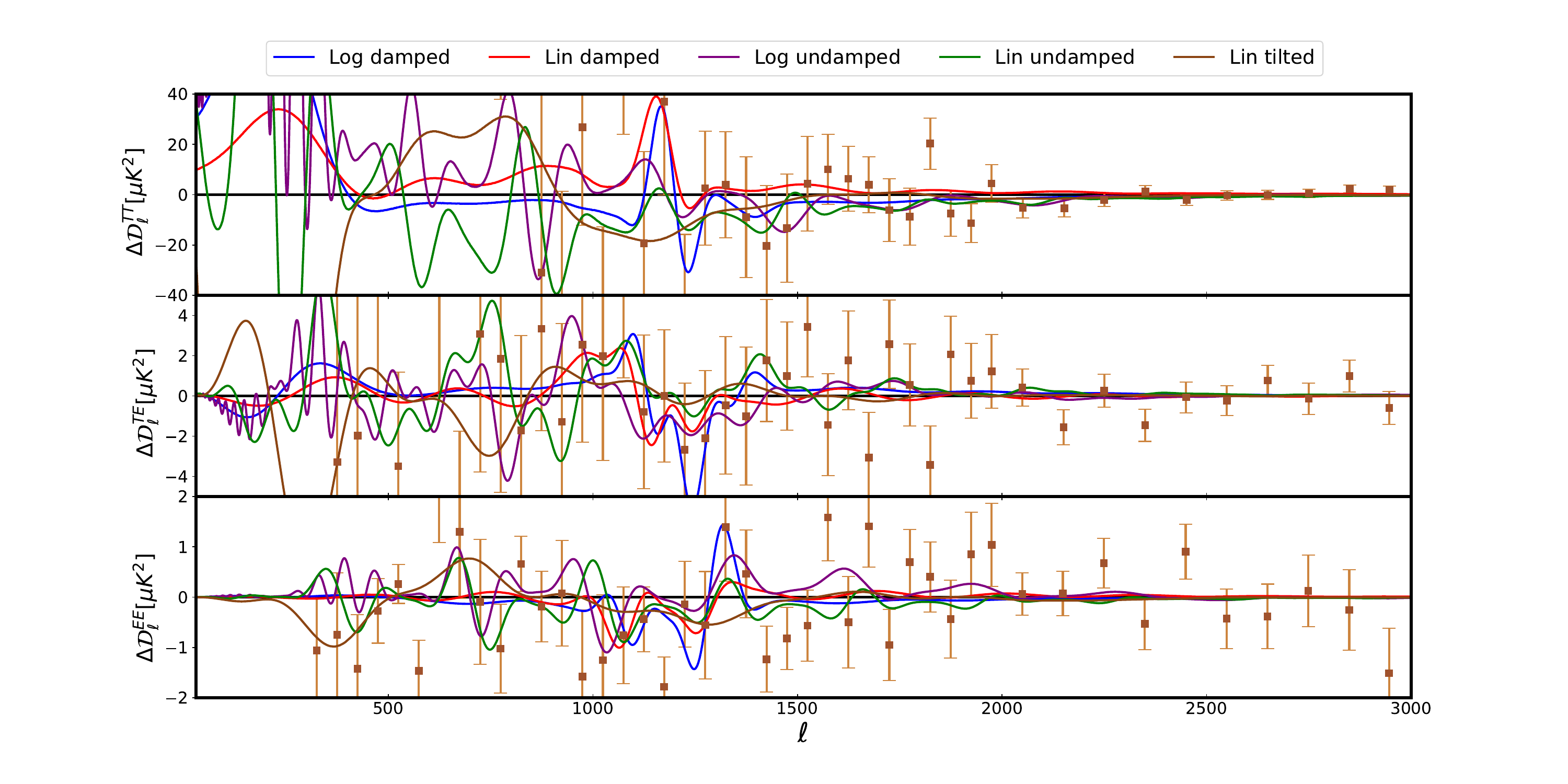}
\includegraphics[width=\textwidth]
{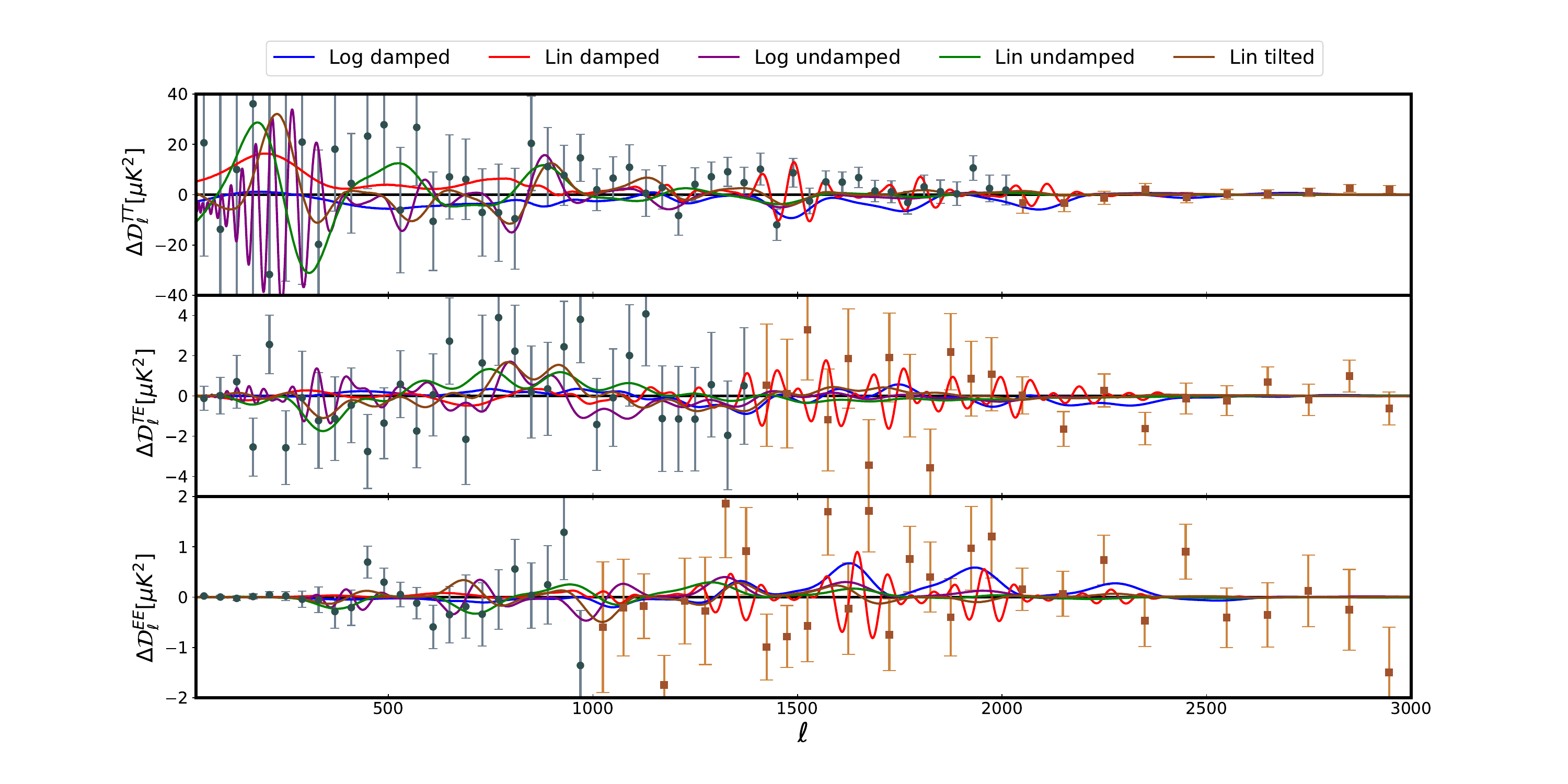} 
\caption{\label{fig:Residual_SPT} Top panel: relative differences of bestfits to the SPT-3G plus $Planck$ 2018 measurement of $\tau$ for the four models considered with respect to the corresponding SPT-3G power law bestfit (see legends for the colors). SPT-3G TT, TE, EE binned residuals (orange data points) to the SPT-3G power law best-fit are also plotted for comparison. Bottom panel: the same for $Planck$ combined with SPT-3G data. 
}
\end{figure*}

In Fig.~\ref{fig:Residual_P18TP} we compare the relative differences of the $Planck$ 2018 TP bestfits for the four models considered with respect to the corresponding $Planck$ 2018 TP power law bestfit with the Plik TT, TE, EE binned residuals. In Fig.~\ref{fig:Residual_SPT} 
we repeat this plot for SPT-3G and the $Planck$ plus SPT-3G combination. These figures show neatly how the bestfits with super-imposed oscillations lead to a better $\Delta \chi^2$ by fitting some of the residuals.


\section{Conclusions}

We have searched for super-imposed oscillations linearly or logarithmically spaced in Fourier wavenumbers in Planck and SPT-3G
2018 temperature and polarization data. We study these super-imposed oscillations either with a constant amplitudes or amplitudes modulated with a Gaussian envelope; to our knowledge it is the first time that the modulated logarithmic oscillation is tested with current CMB data. SPT-3G, with better angular resolution and sensitivity can test primordial features at multipoles beyond the $Planck$ resolution. This is particularly advantageous for features that are at the edge of the $Planck$ angular resolution as the one mimicking the $A_{\rm L}$ effect. Cosmological parameters estimated from SPT-3G 2018 temperature and polarization data~\cite{SPT-3G:2022hvq} in power law $\Lambda$CDM model are completely consistent with $Planck$, and therefore, can be used in combination with $Planck$ to provide a precision test for models
beyond the power law model such as those including features in the primordial power spectrum. See~\cite{SPT-3G:2022hvq,FrancoAbellan:2023gec,Khalife:2023qbu} for other studies of models beyond power law with SPT-3G data.

In this analysis, we use SPT-3G public (binned) data.
We therefore use also $Planck$ binned 2018 data, differently from the search for features performed in Ref.~\cite{Planck:2018inf}, which used unbinned data but fixed nuisance and foreground parameters to their baseline values.
We use PolyChord~\cite{Handley:Polychord} as a sampler given the multimodal nature of the resulting posterior distributions and 
we allow for variation of the foreground and nuisance parameters.

For $Planck$, we find results consistent with those obtained in~\cite{Planck:2018inf} for oscillations with a constant amplitude.
When these oscillations have a Gaussian modulation, linear and logarithmic oscillations produce $\sim - 11.8$ and $\sim - 10.0$ as improvement in $Planck$ 2018 $\Delta \chi^2$, respectively.

We have tested these models with primordial oscillations with SPT-3G 2018 data for the first time. For SPT-3G 2018 data, we have found that all the models studied here improve the fit compared to the power law model, although they are not statistically preferred over the power-law power spectrum. As expected, for $\Lambda$CDM+power law cosmology~\cite{SPT-3G:2022hvq}, SPT-3G is less constraining than $Planck$ 2018 data. Our analysis shows that the frequencies preferred by SPT-3G 2018 data do not coincide with those preferred by $Planck$ 2018 data when the amplitude of the super-imposed oscillations is constant.
However, for a Gaussian modulated amplitude, we find a prominent peak at $\log_{10} \omega_{\rm log} \sim 1.6$ and a broader peak at $\log_{10} \omega_{\rm lin} \gtrsim 1.1$. 
In particular, we find that super-imposed
logarithmic oscillations lead to $\Delta \chi^2 \sim - 12$.
These oscillations do not correspond exactly, but are related to certain features in SPT-3G EE data found by Gaussian regression in~\cite{Calderon:2023obf}
\footnote{Oscillations in $C_\ell^{EE}$ are not necessarily fit by super-imposed oscillations
to the curvature power spectrum which are also imprinted in $C_\ell^{TT}$.}.

We have therefore constrained these models with the combination of $Planck$ and SPT-3G data for the first time.
Since a covariance between $Planck$ and SPT-3G data is not available, we 
consider the main combined data set in which we cut $Planck$ (SPT) above (below) the multipoles at which the
$Planck$ S/N is reached by SPT as stated in~\cite{SPT-3G:2022hvq} and described in section~\ref{sec:data}. This conservative procedure which enforces just one measurement for each multipole bin should overcome the issue of potentially double counting information with a full $Planck$ and SPT-3G combination without a covariance matrix between the two experiments.  By adopting this combination, we find that SPT-3G high-$\ell$ polarization data play an important role for all the models studied.

We find that superimposed oscillations with a constant amplitude provide a 
a $\Delta \chi^2$ larger in the combined data set than in the individual data sets.
The amplitude of these super-imposed oscillations are tighter constrained by $Planck$+SPT-3G 2018 compared to $Planck$ 2018 data.

\begin{figure*}[!htb]
\centering
\includegraphics[width=\textwidth]{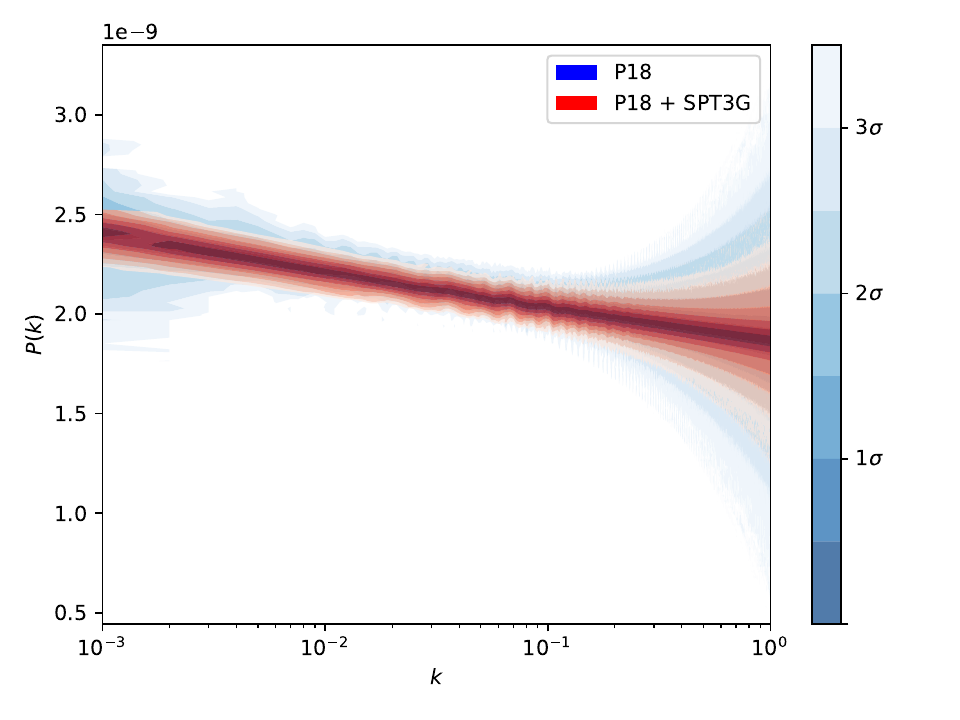} 
\caption{\label{fig:Reconstructed} Reconstructed power spectrum from $Planck$ alone and in combination with SPT-3G data by allowing linear tilted superimposed oscillations described by four additional parameters. 
}
\end{figure*}

{When the amplitude for superimposed linear oscillations is allowed to vary as a power-law in $k$ as in Boundary EFT \cite{Jackson:2013vka}, the $Planck$+SPT-3G 2018 also leads to tighter constraints than the two datasets considered separately. In particular, for the template in Eq. (\ref{eq:undamped-lin}), we find $n_\mathrm{lin} = 0.38 \pm 0.30$ at 68 \% CL with $Planck$+SPT-3G.
The relevance of the SPT-3G data at smaller angular resolution than $Planck$ is nicely illustrated in Fig.~\ref{fig:Reconstructed}, where the reconstructed primordial power spectrum obtained by fgivenx \cite{2018JOSS....3..849H} allowing linear tilted superimposed oscillations is shown for both $Planck$ and $Planck$+SPT-3G data.} 

When the ranges of parameters which provide a better fit independently to $Planck$ and SPT-3G data overlap, we find a larger $\Delta \chi^2$ in the combined $Planck$ SPT-3G data set, 
$\Delta \chi^2 \sim -17.5$ (-14.7) for modulated logarithmic (linear) oscillations. It is interesting to note that SPT-3G data do not prefer $A_L$ larger than 1 \cite{SPT-3G:2022hvq}, but do not disfavour the modulated linear oscillation template which mimics the $A_L$ effect \cite{Planck:2018inf} {yet}. This different response of SPT-3G data to different extensions of the power law model might be due to the number of extra parameters, 1 for $A_L$ and 5 for the modulated oscillation template. {Future experiments dedicated to CMB polarization will be able to disentangle the phenomenological $A_L$ effect from primordial features which mimic it \cite{Antony:2022ert}.}

Our findings could be further improved and extended to higher oscillations frequencies with data which have a finer binning and will be further tested by upcoming CMB temperature and polarization measurements at high multipoles by ongoing and future ground experiments, such as ACT \cite{AtacamaCosmologyTelescope:2025blo}, SPT \cite{SPT-3G:2025bzu} and Simons Observatory~\cite{SO}. 

\begin{acknowledgments}
The authors acknowledge the use of computational resources
at the Institute of Mathematical Science’s High Performance Computing facility [Nandadevi] and of CNAF HPC cluster in Bologna.  We would like to thank Jan Hamann for useful comments on the manuscript. FF and DKH would like to thank Karim Benabed and Silvia Galli for discussions on SPT-3G 2018 likelihood. AA is supported by an appointment to the JRG Program at the Asia Pacific Center for Theoretical Physics through the Science and Technology Promotion Fund and Lottery Fund of the Korean Government, and was also supported by the Korean Local Governments in Gyeongsangbuk-do Province and Pohang City. AA also acknowledges support from the NRF of Korea (Grant No. NRF-2022R1F1A1061590) funded by the Korean Government (MSIT). AA, FF, DKH acknowledge travel support through the India-Italy "RELIC - Reconstructing Early and Late events In Cosmology" mobility program. 
FF would like to thank IMSc for warm hospitality for the December 2022 visit when this work started. 
FF, DP acknowledge financial support from the contract by the agreement n. 2020-9-HH.0 ASI-UniRM2 ``Partecipazione italiana alla fase A della missione LiteBIRD". FF, DKH, DP acknowledge financial support from Progetti di Astrofisica Fondamentale INAF 2023. DKH would like to acknowledge the support from the Indo-French Centre for the Promotion of Advanced Research – CEFIPRA grant no. 6704-4.
AS would like to acknowledge the support by National Research Foundation of Korea NRF-2021M3F7A1082056, and the support of the Korea Institute for Advanced Study (KIAS) grant funded by the government of Korea. 
\end{acknowledgments}




\bibliography{main}
\end{document}